\documentclass[acmtog, authorversion]{acmart}

\AtBeginDocument{%
  \providecommand\BibTeX{{%
    \normalfont B\kern-0.5em{\scshape i\kern-0.25em b}\kern-0.8em\TeX}}}

\setcopyright{acmcopyright}
\copyrightyear{2022}
\acmYear{2022}
\acmDOI{10.1145/1122445.1122456}

\acmConference[]{ ACM Symposium on the Science of Security (HOTSOS)}{April 03, 2023}{}
\acmBooktitle{ ACM Symposium on the Science of Security (HOTSOS),
  April 03, 2023}
\acmPrice{15.00}
\acmISBN{978-1-4503-XXXX-X/18/06}
 \usepackage{multirow}
 \usepackage[normalem]{ulem}
 \useunder{\uline}{\ul}{}
\usepackage{lineno}
\usepackage{caption}
\usepackage{subcaption}
\modulolinenumbers[5]
\usepackage{graphicx}
\usepackage{booktabs}

\usepackage{amssymb,amsmath,nccmath}
\usepackage{cclicenses}
\usepackage{makecell}
\usepackage{lscape,array}
\usepackage[thin, , thinc]{esdiff}
\usepackage{subcaption}
\usepackage{caption}
\usepackage{framed}  
\usepackage[font=small,skip=0pt]{caption}
\usepackage{xcolor}
\usepackage{tabularx}
\newcommand{\G}{\mathcal{G}}

\begin{document}

\title{Sequential Graph Neural Networks for Source Code Vulnerability Identification}

\author{Ammar~Ahmed}
\email{ahme0599@umn.edu}
\affiliation{%
  \institution{University of Minnesota}
  \city{Minneapolis}
  \state{Minnesota}
  \country{USA}
}

\author{Anwar~Said}
\email{anwar.said@vanderbilt.edu}
\affiliation{%
  \institution{Vanderbilt University}
  \city{Nashville}
  \state{TN}
  \country{USA}}
\email{anwar.said@vanderbilt.edu}

\author{Mudassir~Shabbir}
\email{mudassir.shabbir@vanderbilt.edu}
\affiliation{%
  \institution{Vanderbilt University}
  \city{Nashville}
  \state{TN}
  \country{USA}
}

\author{Xenofon~Koutsoukos}
\email{xenofon.koutsoukos@vanderbilt.edu}
\affiliation{%
 \institution{Vanderbilt University}
 \city{Nashville}
 \state{TN}
 \country{USA}}


\begin{abstract}

Vulnerability identification constitutes a task of high importance for cyber security. It is quite helpful for locating and fixing vulnerable functions in large applications. However, this task is rather challenging owing to the absence of reliable and adequately managed datasets and learning models. Existing solutions typically rely on human expertise to annotate datasets or specify features, which is prone to error. In addition, the learning models have a high rate of false positives. To bridge this gap, in this paper, we present a properly curated C/C++ source code vulnerability dataset, denoted as CVEFunctionGraphEmbeddings (CVEFGE), to aid in developing models. CVEFGE is automatically crawled from the CVE database, which contains authentic and publicly disclosed source code vulnerabilities. We also propose a learning framework based on graph neural networks, denoted SEquential Graph Neural Network (SEGNN) for learning a large number of code semantic representations. SEGNN consists of a sequential learning module, graph convolution, pooling, and fully connected layers. Our evaluations on two datasets and four baseline methods in a graph classification setting demonstrate state-of-the-art results.

\end{abstract}



\keywords{Graph Neural Networks, Vulnerability Dataset Curation, Software Security, Vulnerability Identification}

\maketitle

\section{Introduction}
\label{introduction}


The collaborative nature of open-source software development has exponentially expanded its popularity in recent years. The open-source software sector is worth more than $\$21.7$ billion and is predicted to hit $\$50$ billion by 2026. As the number of open-source projects grows, so do the vulnerabilities in the software sector. The total number of vulnerabilities registered in the Common Vulnerabilities and Exposures (CVE) database climbed from $4600$ in 2010 to approximately $164000$ in 2021, demonstrating this trend \cite{CVEdatabase}. These vulnerabilities, which are frequently caused by unsecured source code, enable cyberattacks and may inflict significant social and financial impact.

Automatic vulnerability detection at an early stage is a widely established strategy for avoiding cyber attacks. However, this task is quite challenging due to a variety of source codes and cyber threats, lack of datasets, and the need to include human experts \cite{li2018vuldeepecker}. To date, a variety of techniques for automatically detecting software vulnerabilities have been introduced. On the one hand, traditional approaches such as static analyzers \cite{xu2017spain,chandramohan2016bingo}, dynamic analyzers, and symbolic execution methods \cite{li2019cerebro,chen2018hawkeye,wang2019superion,li2017steelix,xue2018accurate} are prevalent. However, these solutions have three key drawbacks: a high false-negative rate, extensive human labor, and the paucity of labeled data. In contrast, Machine Learning (ML) techniques have been tested more recently as a complementary approach to circumvent the limitations of the current methodologies \cite{zhou2019devign,neuhaus2007predicting,shin2010evaluating,harer2018automated}. Though ML methods have demonstrated promising results given the availability of a few labeled datasets, the results are not yet applicable to a real-world setting and new methods are required to bridge the gap \cite{said2021netki,hanif2021rise}.


Recently, few attempts have been made to automate feature engineering using deep learning frameworks in order to avoid the labor-intensive task of human experts \cite{zhou2019devign,russell2018automated}. The authors in \cite{zhou2019devign} employ Graph Neural Networks (GNNs) for automatic vulnerability identification in source code. However, this framework doesn't allow to consider graphs of different sizes. Moreover, the previous methods either considered the source code as a flat sequence, employing some incomplete information and the control flow graph with \textit{word2vec} embeddings as node features. Also, they consider constructing datasets using human experts, which can be prone to error. Source code, on the other hand, has a more diverse structure, with representations such as Control Flow Graph (CFG) and Abstract Syntax Tree (AST), among others. Nonetheless, the vulnerabilities are frequently subtle, necessitating a thorough analysis from several semantic perspectives. Simply considering the flat sequence or conducting some partial feature engineering would not be sufficient to distinguish vulnerable from non-vulnerable code. A similar challenge can be seen when generating graphical representations of source code such as CFG and AST. It's difficult to describe vulnerable codes in graph structure in a way that distinguishes them from non-vulnerable codes. The generation of node embeddings from source code is another major problem. Although word2vec embeddings created from the source code are used in numerous techniques \cite{zhou2019devign,wang2020combining,liu2021combining,said2021dgsd}. However, existing approaches continue to have a significant false-negative rate due to poor intermediate representations of source code.

\begin{figure*}[!t]
    \centering
    \includegraphics[width = 0.85\textwidth]{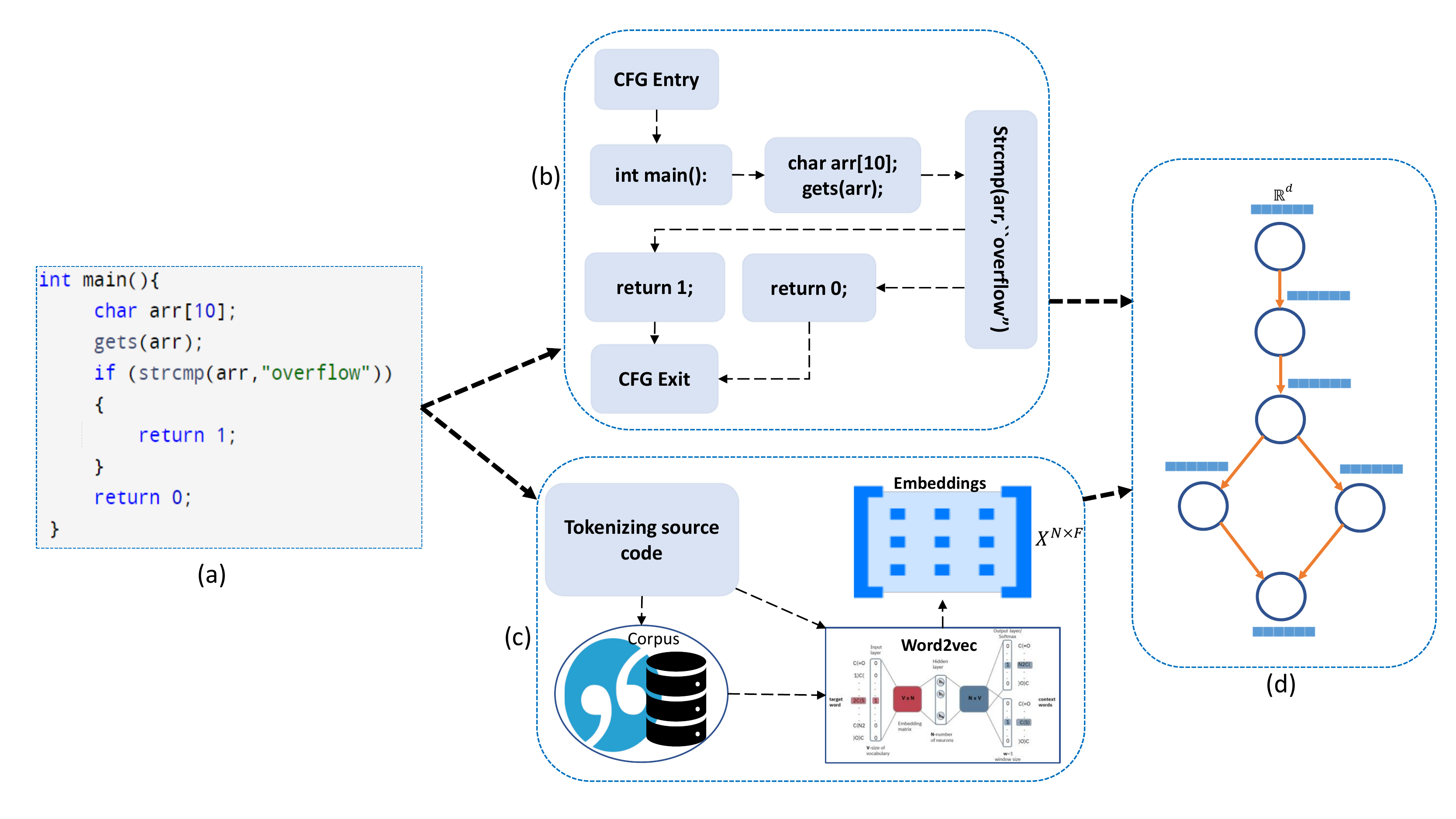}
    \caption{Illustration of creating CFGs, embeddings, and the resulting attributed graph from the source code is shown.  (a) shows the source code, whereas (b) shows the corresponding CFG. The embeddings module is demonstrated in (c), which includes source code tokenization, corpus creation, and the word2vec model. The resultant graph is presented in (c). }
    \label{fig:example}
\end{figure*}


To address the aforementioned limitations, we offer two key contributions in this work. First, we collected a vulnerability dataset  we denoted as CVEFunctionGraphEmbeddings (CVEFGE) from the CVE database, which contains more authentic real-world vulnerabilities. Second, we introduce a novel framework, we named SEquential Graph Neural Network (SEGNN) for early-stage vulnerability detection based on GNNs. As inputs to SEGNN, we consider the CFG representation of the source code and word2vec embeddings as node features and train an end-to-end GNNs framework. We illustrate the different steps involved in the graph construction in Figure \ref{fig:example}. Unlike existing methods, our framework is based on attributed graphs and employs graph attention mechanisms, as well as a sequential learning module, to learn on a diverse set of code semantic representations. We summarize the key contributions of this study as follows.  

\begin{itemize}
    \item We crawled the CVE database and plan to publish a comprehensive vulnerability dataset (CVEFGE) with all CVE features as well as graphical representations in the form of CFGs and learned embeddings. 
    \item We present a GNN architecture with sequential learning modules and an attention and pooling technique for the early detection of source code vulnerabilities.
    \item We compared the performance of several learning techniques on our dataset to that of an existing dataset and demonstrate that the performance of the learning models is up to 20\% better on our dataset. 
    \item On both datasets, we compare the performance of the proposed framework to four baselines and demonstrate superior results.
\end{itemize}



The following is the breakdown of the paper's structure. Section 2 reviews the relevant literature with a focus on ML approaches to vulnerability identification. Section 3 presents the curation of the vulnerability dataset. The proposed framework is presented in Section 4, and the experimental setup and findings are presented in Section 5. Finally, Section 6 concludes the paper by discussing some promising future directions.

\section{Related Work}
\label{related-work}

Identifying vulnerabilities at an early stage is essential for any software system to be reliable. In the last few years, several attempts have been made to advance the field \cite{lin2020software}. We classify these methods into three different categories and discuss them separately in the following sections.

\subsection{Graph-based Feature Representations}


Exploring the structure of a program using graph-based representations such as CFG, AST, PDGs (Program Dependence Graphs), and DDGs (Data Dependence Graphs) is a frequently used approach \cite{lin2017poster,wang2016automatically}. CFG is a graphical representation of computations performed during the execution of a program made up of basic blocks, a starting block, and an ending block. Starting and ending blocks represented the start and end of the program's execution, whereas basic blocks represented a series of straight lines or branch-free code. PDGs and DDGs give more detailed and data-dependency-based representations of the program, while AST gives a more natural and abstract representation of the program's grammar \cite{zhou2019devign}. 

Wang et al., \cite{wang2016automatically} investigated the use of an AST representation to examine the Java program structure automatically to identify early-stage vulnerabilities. In the AST, three different types of nodes were maintained. First, some nodes represent the generation of a class instance and the invocation of a function. Second, the declaration nodes, and finally the control-flow nodes. To reduce noise from the data, token sequences were constructed from the AST nodes, and then a distance-based technique \cite{navarro2001guided} was used. The $k-$nearest neighbors method was then used to calculate distances between sequences, and nodes were labeled as noise if they had labels that were opposite to their neighbors. The authors of \cite{lin2017poster} developed a function-level AST-based strategy for learning feature representations that employs Bi-LSTM. To identify vulnerable functions, the authors used a software complexity metric and depth-first traversal to generate sequences from AST. Following that, the sequences were transformed into word embeddings and put into the Bi-LSTM model for learning. A similar approach was used in \cite{lin2018cross} to train the Bi-LSTM model, which used token sequences generated from AST and word2vec embeddings. In \cite{dam2017automatic}, the authors introduced a sequence to sequence the LSTM network for detecting file-level vulnerabilities. Sequences were first constructed from program tokens, and then those sequences were utilized to generate vector embeddings, which were then input into the LSTM model. Further, \cite{li2021sysevr} proposed a somewhat more advanced approach for identifying different types of vulnerabilities using Bi-LSTM and Bi-GRU. The authors used CFGs to transform syntax-based vulnerability candidates into semantic-based vulnerability candidates to enrich the embeddings with syntax and semantic information. DFGs were also created, and word2vec embeddings were used to produce the functions' final embeddings.

The authors in \cite{zhou2019devign} have recently introduced a more comprehensive representation of source code that includes CFGs, ASTs, DFGs, and NCS (Natural Code Sequence) representations. To learn function-level embeddings, a GNNs framework comprising graph embedding layers, a gated graph recurrent layer, and a convolution module was presented. Vulnerability identification is treated as a graph classification problem in this paradigm. This work also included the creation of function-level labeled datasets tagged by human experts and shows encouraging results on vulnerability identification tasks as compared to the existing methods. The work in \cite{wang2020combining} takes a similar method, constructing intermediary representations (relational graphs) using AST, program control, and dependent graphs, and then using a GNNs model to identify vulnerabilities.

\subsection{Sequence-Based Feature Representations}

Sequence-based methods usually follow the Neural Networks (RNNs) to learn feature representations that include the sequence of function calls, system execution trace, sequence of statements, etc. In the following, we provide a brief overview of these methods.

The authors in \cite{grieco2016toward}  used feature sets taken from both static and dynamic analyzers to identify memory corruption vulnerabilities. The essential premise was that the analyzers' call sequence reveals patterns that can be used to discover memory corruption. To extract embeddings from the input sequences, static and dynamic analyzer call sequences are obtained first, and then N-gram and word2vec models are utilized. For classification, the resultant embeddings were fed to random forest and neural network classifiers. Another method for discovering buffer and resource management error vulnerabilities is to employ a ``Code Gadget" representation of the application \cite{li2018vuldeepecker}. The code gadget is a sequence of statements representing variable flows and data dependencies made up of several consecutive lines of code that are semantically tied to each other. After that, Code Gadget was transformed to Word2Vec embedding, and Bi-LSTM was used to find vulnerabilities. In \cite{li2021vuldeelocator}, the authors extended \cite{li2018vuldeepecker} and added flow-level information to the Code Gadget. Furthermore, a code attention approach was presented to extract more localized information from statements. The datasets have also been made publicly available by the aforementioned two works \cite{li2021vuldeelocator,li2018vuldeepecker}.

\subsection{Text-Based Feature Representations}

Text-based representation approaches treat source code as text and generate meaningful representations using various text-based methods. In \cite{peng2015building}, the authors tokenized Java source code and used the N-gram model to extract feature representations. Wilcoxon rank-sum \cite{wilcoxon1992individual} was used to reduce the dimensions of the feature vectors to avoid high dimensionality. The authors then used a neural network to make predictions. In \cite{russell2018automated} the authors collected a dataset consisting of 12 million source code functions and extended \cite{peng2015building} to detect vulnerabilities in $C/C++$ source code. More recently, Variational Autoencoder was applied to the embeddings of sequences of machine instructions for detecting vulnerabilities at the binary level \cite{kingma2013auto}.

For early-stage vulnerability detection, we use a novel approach to extract syntactical and semantic information from source code. Unlike previous works that use word2vec embeddings, we use learnable embeddings for node features and subsequently offer a GNNs-based vulnerability detection approach. This enables us to extract fine-grained information from source code and include it in the learning process. In addition, we present extensive experimental results based on word2vec embeddings incorporated with the existing models. 
\section{Dataset Curation}
\label{dataset-collection}


Several efforts have been made in recent years to acquire relevant datasets for vulnerability identification using machine learning \cite{zhou2019devign,ponta2019manually,fan2020ac}. The authors in \cite{fan2020ac} have gone to tremendous lengths to crawl the CVE database and collect vulnerable source code denoted as Big-Vul from a variety of open-source repositories. The Big-Vul dataset contains the vulnerable source codes, as well as the fixed versions and other descriptive information. However, the present dataset has only $11823$ vulnerable functions, including graphs of size $10$. After applying a filter with a graph of size $10$, only $8000$ occurrences remain. In addition, it comprises around $91$ unique categories of vulnerabilities, the bulk of which have a tiny number of instances, making it challenging for ML models to generalize. In addition, Big-Vul lacks the graphical representation of the functions, necessitating considerable time and effort to prepare it for the graph machine learning tasks. To overcome these limitations, we crawled the CVE database and gathered CVE entries between $2002$ and $2021$. We parse the web pages containing CVE information for each entry, year by year. Then, we restricted our search to just those entries that led to publicly available repositories. Using the given URLs, we extracted the vulnerable functions and their relevant information from the CVE database and included them in our dataset, CVEFGE. 

\begin{figure}
\centering
  
  \includegraphics[width=\linewidth]{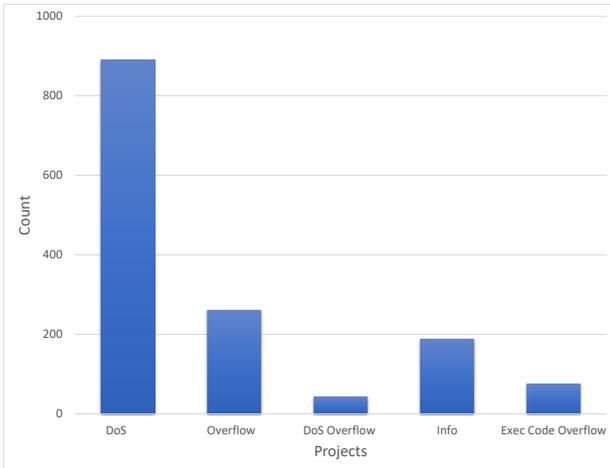}
  \caption{Count of top 5 vulnerabilities in CVEFGE.}
  \label{fig:vul-type}
\end{figure}

\begin{figure}
  \centering
  \includegraphics[width=\linewidth]{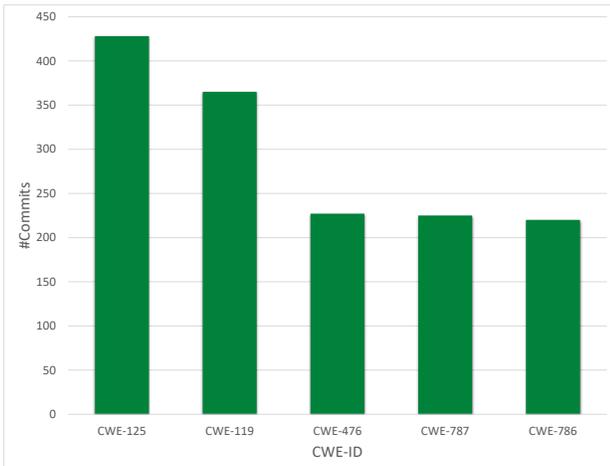}
  \caption{Number of commits against top 5 CWE in CVEFGE}
  \label{fig:num-commits}
\end{figure}


Our data collecting procedure closely follows the Big-Vul data crawling strategy and involves the following steps. a) Utilizing the CVE database's yearly records, we crawled the web pages of each CVE entry and collected descriptive information including the CVE ID, CVE summary, CWE ID, publish date, and vulnerability type. b) We search for CVE IDs with accessible Git repository URLs and automatically retrieve their links. c) We retrieve the files containing the vulnerable functions' source code. d) Since each file may include many source code functions, we divided the files and extracted both vulnerable and non-vulnerable routines. e) Using the Joern tool, we generate their respective CFGs. f) Using the suggested technique, we added the learned embeddings to each graph. We save CFGs in the ``dot" format, which is readily accessible by graph tools such as Python's networkx. We report the descriptive statistics for the CVEFGE dataset in Table \ref{tab:new-big-vul} and the graph dataset's features in Table \ref{tab:des-stats}.

\begin{table}[!htb]
    \centering
    \begin{tabular}{|ll|l|l|l|}
    \hline
    \multicolumn{2}{|l|}{}                                                   & \textbf{All} & \textbf{Positive} & \textbf{Negative} \\ \hline
    \multicolumn{1}{|l|}{\multirow{2}{*}{\textbf{Graphs}}}      &  \textit{Complete}  &$32614$      &     $9190$           &        $23424$       \\ \cline{2-5} 
    \multicolumn{1}{|l|}{}                                & \textit{Balanced}     & $18390$     &      $9190$          &       $9200$   \\ \cline{2-5} \hline
    
    \multicolumn{5}{|l|}{\textbf{}}  \\ \hline  
    \multicolumn{1}{|l|}{\multirow{3}{*}{\textbf{Nodes}}} & \textit{Max}     & 4755         & 4755               & 4749              \\ \cline{2-5} 
    \multicolumn{1}{|l|}{}                                & \textit{Min}     & 11           & 11                & 11                 \\ \cline{2-5} 
    \multicolumn{1}{|l|}{}                                & \textit{Average} & 81.91         & 59.73               & 138.45         \\ \hline

    \multicolumn{5}{|l|}{\textbf{}}                                     \\ \hline
    \multicolumn{1}{|l|}{\multirow{3}{*}{\textbf{Edges}}} & \textit{Max}     & 6264         &  6264            & 5578              \\ \cline{2-5} 
    \multicolumn{1}{|l|}{}                                & \textit{Min}     & 10            & 10               & 10                 \\ \cline{2-5}
    \multicolumn{1}{|l|}{}                                & \textit{Average} & 92.25           &  66.26              & 158.49                \\ \hline
    \multicolumn{5}{|l|}{\textbf{}}                                     \\ \hline

    \multicolumn{1}{|l|}{\multirow{3}{*}{\textbf{Density}}} & \textit{Max}     & 0.2836        & 0.2836             & 0.1859              \\ \cline{2-5} 
    \multicolumn{1}{|l|}{}                                & \textit{Min}     & 0.0002            & 0.0002                 & 0.0002                 \\ \cline{2-5}
    \multicolumn{1}{|l|}{}                                & \textit{Average} & 0.0357           & 0.0398               & 0.0251               \\ \hline
    
    \end{tabular}
    \caption{Characteristics of the CVEFGE dataset.} 
    \label{tab:new-big-vul}
\end{table}

\begin{table}
  
    \centering
    \begin{tabular}{|l|c|c|} \hline
         \textbf{Measurements}& \textbf{Unique Count} &  \textbf{Total Count} \\ \hline
        CWE ID & $110$& $3030$ \\ \hline
         CVE ID& $3226$& $3226$ \\ \hline
        Commits &$3438$& $6877$  \\ \hline
        Number of Projects & - & $517$  \\ \hline
        Vulnerability Types & - & $50$  \\ \hline
        Vulnerable Functions & - & $9190$ \\ \hline
        Non Vulnerable Functions & - &  $23424$ \\ \hline
    \end{tabular}
    \caption{Descriptive stats of CVEFGE dataset}
    \label{tab:des-stats}
\end{table}

In CVEFGE, we identified DoS, Overflow, DoS Overflow, Info, and Exec Code Overflow as the top five vulnerabilities. Their stats are shown in Figure~\ref{fig:vul-type}. We also show CWE with the most commits in Figure \ref{fig:num-commits}.


\section{Vulnerability Identification Framework}
\label{methodology}


Previous studies have shown that graphical representations such as AST, CFGs, and Code Property Graph (CPG) are among the most effective ways for representing the syntactical and semantic information of source code \cite{zhou2019devign,wang2020combining}. Control flow and semantic information are critical for understanding different types of source code vulnerabilities. Moreover, the combination of syntactic information and graphical representations contributes quite successfully, as shown by a recent study \cite{zhou2019devign}. Motivated by these intriguing results, we focus our efforts in this study on constructing a machine learning framework that can learn the program's syntax and semantics and generate predictions concurrently with the curation of the dataset. In the following sections, we first present the problem setup, followed by the specifics of our proposed graph learning approach for identifying source code vulnerabilities.

\begin{figure*}
    \centering
    \includegraphics[width = 0.95\textwidth]{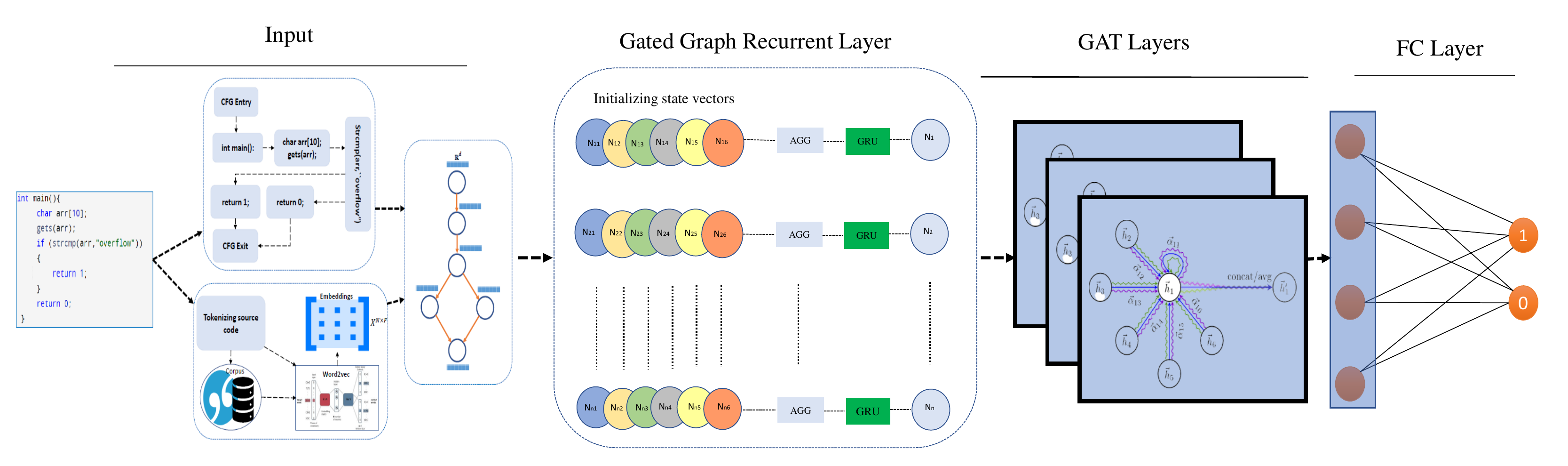}
    \caption{Architecture of the proposed framework, SEGNN. }
    \label{fig:architecture}
\end{figure*}


\subsection{Problem Setup}

We consider identifying \textit{function level} vulnerabilities, which form a binary classification problem. Several previous works have addressed the issue at the application or file level \cite{li2018vuldeepecker,dam2017automatic}. However, recent studies indicate that function-level analysis provides a finer level of granularity in vulnerability detection \cite{zhou2019devign,wang2020combining}. Lets have $(g_i, y_i)|g_i \in \G, y_i \in \mathcal{Y}, i \in \{1,2,\ldots, n\}$ where $\G$ denotes a set of source code functions represented in the form of CFGs. $n$ represents the number of instances, whereas $\mathcal{Y} \in \{0,1\}^n$ indicates the label set representing 1 for vulnerable instances and 0 otherwise. Let $\mathcal{X}^{V \times F}$ denotes nodes' feature matrix associated with each graph $g_i$ where $V$ indicates the total number of nodes in $g_i$ and $F$ is the size of the embedding vector $h_i$ (see section \ref{node-embedding-layer}). Lets $A$ denotes the adjacency matrix. We define a function of the source code as $g_i(V, \mathcal{X}, A, y) \in \G$ and the goal is to learn a function $f:\G \rightarrow \mathcal{Y}$ that predicts a label $y_i$ for an unseen instance $g_i$. 


\subsection{SEquential Graph Neural Network (SEGNN)}
\label{GNNs-model}

The proposed framework comprises primarily of four modules: the embedding layer, the sequential learning module, the graph Convolution with attention layer, and the graph pooling layer. We provide the illustration of the proposed framework in Figure \ref{fig:architecture}. Following is a breakdown of each of these modules. 

\subsubsection{Embedding Layer}
\label{node-embedding-layer}
The performance of ML models for vulnerability identification relies heavily on the extraction of fine-grained information from the source code. However, the modification in the source code is fairly subtle sometimes, such as null pointer dereferencing and Denial of Service (DoS) attacks, making it extremely difficult to spot without a deep dive into miner-level details. Therefore, creating expressive node embeddings is essential to the model's performance \cite{zhou2019devign,wang2020combining,liu2021combining}. Thus, we consider Word2Vec embeddings as node features along with CFGs as input to our model.


\textbf{Control Flow Graph (CFG):} CFG is used to simulate the control flow between the different components of a program. It consists of a beginning block, an ending block, and many basic blocks. Beginning and ending blocks represent the beginning and end of a program's execution, respectively, while basic blocks represent the units of a branch-free code that runs in sequence (statements and conditions). The control flow establishes directed edges between CFG blocks, with each block representing a node. 

\textbf{Node Features:} The node attributes have a significant impact on the performance of any graph machine learning model. We use both the basic-block (node level) and whole-function (graph level) source codes. To create these embeddings, we use a source code tokenizer, build a corpus, and train a word2vec model. Since source codes and their types are available for each node of the network, we create corresponding word2vec embeddings for each node. Then, using the same trained model, we produce graph-level embeddings (feature vector for the whole graph) and concatenate them with node embeddings. These two embeddings are regarded as the initial node encoding $x_v$. Our embedding layer is defined as follows:

\begin{equation}
    \label{eq:embedding}
    g_i(V, \mathcal{X}, A, y) = \mathrm{EMBEDDINGS}(f_i), \forall i= \{1 \ldots, n\}
\end{equation}
Where $f_i$ indicates the input function of the source code. We show an example of the embedding layer (generation of CFGS and Embeddings and attributed graph in Figure \ref{fig:example}.

\subsubsection{Sequential Learning Module}
\label{GGRN}

Sequential information in source code has shown to be quite useful for discovering vulnerabilities. Learning a sequence guarantees that both the semantics and the order of actions among each other are captured. To this end, we use Gated Graph Recurrent Networks (GGRN) \cite{li2015gated}, a form of Recurrent GNNs, to learn node representations from both a local and global perspective. Several other GNN architectures, such as graph attention network (GAT) \cite{velivckovic2017graph}, graphSAGE \cite{hamilton2017inductive}, and graph convolutional networks \cite{kipf2016semi}, can be considered for this task, but GGRN is better suited for learning sequences and graph topology at a deeper level.

We initialize the state vector with zeros $h_i \in \mathbb{R}^z $ where $z \geq F$ for each node $v_i \in V$ in $g_i$. The initial node embeddings $x_i$ is copied into the first dimensions when $h_i$ is initialized. By that we have $h_i^1 = [x_i^\top, \boldsymbol{0}]^\top$ for the initial node embeddings. Then the basic propagation model is defined as follows: 

\begin{equation}
    \label{eq:propagation}
     \boldsymbol{a}_v^{(t-1)}  = W_{v:}^\top \left[h_1^{(t-1)\top}, \ldots, h_V^{(t-1)\top} \right]^\top + b
\end{equation}

Where $W$ is the weight matrix and $b$ is the bias. With equation \ref{eq:propagation}, we aggregate information from all neighbor nodes to calculate a new state for a node $v$. Afterward, we use Gate Recurrent Unit (GRU) to extract information of all the states to get the embeddings $h_v$ for the current node. 

\begin{equation}
    \label{eq:GRU}
    h_v^{(t)} = \text{GRU} \left( h_v^{(t-1)},  a_v^{(t-1)}\right)
\end{equation}
The final state vectors are obtained as an embedding matrix after $T$ iterations of the aforesaid propagation technique.

\subsubsection{Graph Convolution with Attention}
\label{gcn-layers}

The second module of our architecture is comprised of three blocks of graph convolution layers accompanied by an attention mechanism. The attention mechanism is used in the recursive neighborhood diffusion process in order to integrate minute modifications in the source code to the latent representation \cite{velivckovic2017graph}. The overall architecture is defined as follows:

\begin{equation}
\label{eq:attention}
    e_{ij} = g(\boldsymbol{W}h_i, \boldsymbol{W}h_j)
\end{equation}
$\boldsymbol{W^{F^\prime \times F}}$ is the weight matrix, $h_i$ is the latent representation of node $i$ and $g:\mathbb{R}^{F^\prime} \times  \mathbb{R}^{F^\prime} \rightarrow \mathbb{R}$ is a single-layer feed forward neural network. $e_{ij}$ indicates the importance of node $j'$s features to node $i$. $e_{ij}$ is further normalized through Softmax to make it easily comparable across different nodes.   

\begin{equation}
\label{eq:softmax}
    \alpha_{ij} = \text{softmax}(e_{ij}) 
\end{equation}

With the above attention layer, the convolution is defined as follows:

\begin{equation}
    \label{convolution}
    h_i = \sigma \left( \sum_{j \in \mathcal{N}_i} \alpha_{ij}\mathbf{W}h_j \right)
\end{equation}
    
Where $\mathcal{N}_i$ is the set node $i$'s neighbors.

\subsubsection{Pooling Module}
\label{pooling-module}


Typically, the pooling module is the last module in GNN models that combines the node representation into a single feature vector representing the complete graph. Various methods of pooling, including mean-pooling, max-pooling, and add-pooling, may be used to provide the appropriate graph-level properties. We used global-max-pooling followed by a linear layer in our experimental setup.

\section{Experiments}
\label{evaluation}
In this section, we present the evaluation of our proposed model and dataset against four baselines and one existing dataset.
\begin{figure*}[!t]
\centering
\begin{subfigure}{.45\textwidth}
  \centering
  \includegraphics[width=\linewidth]{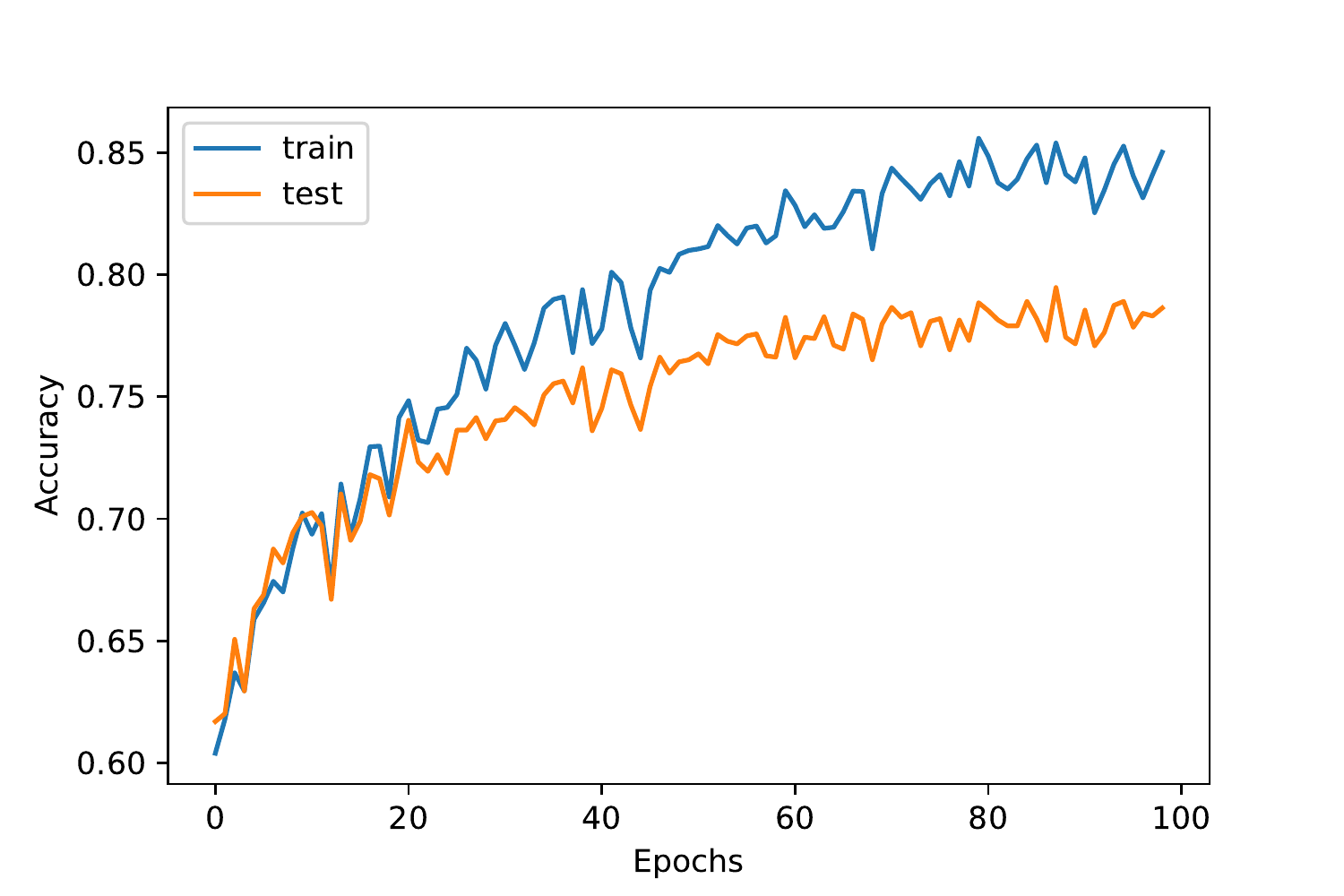}
  \caption{CVEFGE dataset}
  \label{fig:curves-extended}
\end{subfigure}%
\begin{subfigure}{.45\textwidth}
  \centering
  \includegraphics[width=\linewidth]{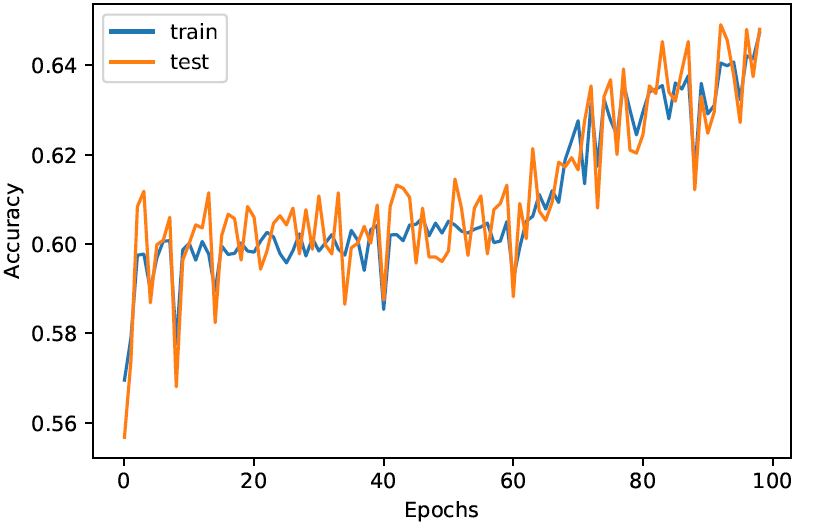}s
  \caption{Big-Vul dataset}
  \label{fig:curves-big-vul}
\end{subfigure}
\caption{Training and Testing Curves of the proposed framework}
\label{fig:curves}
\end{figure*}
\subsection{Experimental Setup:}

\textbf{Setting:} The proposed architecture is comprised of one gated graph convolution layer, three graph attention layers, a max-pooling layer, and a linear layer. We set the output channel to 200, the number of layers to 3, and the aggregator function for the convolution layer of the gated graph to ``add". We employ 64, 64, 32, and 16 neurons in the attention layer and 16 neurons in the final linear layer. All experiments were conducted on an Intel Core i9 system with 64 GB of RAM and a NVIDIA GeForce 6GB GPU, with the batch size set to 128 and the learning rate set to $0.001$. We use a train-test split of $80:20$ and accuracy as a performance metric for the model.

\textbf{Baselines:} We consider MEWISPool \cite{nouranizadeh2021maximum}, GraphSAGE \cite{hamilton2017inductive}, GAT \cite{velivckovic2017graph} and DGSD \cite{said2021dgsd} frameworks for comparison. MEWISPool is a recently proposed graph classification framework based on the GNN paradigm. It provides a novel pooling layer for graphs that optimizes the information gain between the input and pooled graphs. DGSD is a statistical graph descriptor that uses graph statistics to construct a graph representation in a single step. GAT and GraphSAGE, are well-known graph neural network models.


\textbf{Data Preprocessing:}

Vulnerability datasets are typically provided as source code; therefore, further preprocessing is required to prepare the data for the model. To produce CFGs, we use the Joern tool\footnote{\url{https://joern.io/}}. Joern is a code analysis tool with other functions than the generation of CFGs, ASTs, and CPGs. It provides a JSON representation of each CFG, with labels and source code for every CFG block. To generate node features, we first tokenize the whole code and construct a corpus from which the word2vec model will be trained. Next, we train the word2vec model using the source code associated with each node to generate node features using the trained word2vec model. We also eliminated the graph with the size of $< = 10$ and converted the data to torch geometric\footnote{url{https://pytorch-geometric.readthedocs.io/en/latest/}}.

\textbf{Datasets: } 


In our experimental setting, we investigate the balanced versions of two datasets: Big-Vul \cite{fan2020ac} and the proposed CVEFGE. The statistics for both datasets are shown in Table \ref{tab:big-vul} and Table \ref{tab:new-big-vul}, respectively. CVEFGE has a total of $46,097$ instances, of which $10,883$ are vulnerable and $35,214$ are not vulnerable. After excluding graphs with a size of less than 10, we are left with $32,614$ instances, of which $9,190$ are vulnerable and $23,424$ are not. We randomly selected $9200$ non-vulnerable instances to balance the dataset.

\begin{table}[!htb]
\centering
\begin{tabular}{|ll|l|l|l|}
\hline
\multicolumn{2}{|l|}{}                                                   & \textbf{All} & \textbf{Positive} & \textbf{Negative} \\ \hline

\multicolumn{1}{|l|}{\multirow{3}{*}{\textbf{Nodes}}} & \textit{Max}     & 9011         & 9011              & 3766              \\ \cline{2-5} 
\multicolumn{1}{|l|}{}                                & \textit{Min}     &11            & 11                 & 11                 \\ \cline{2-5} 
\multicolumn{1}{|l|}{}                                & \textit{Average} & 98           & 132              & 54                \\ \hline

\multicolumn{5}{|l|}{\textbf{}}                                     \\ \hline
\multicolumn{1}{|l|}{\multirow{3}{*}{\textbf{Edges}}} & \textit{Max}     & 11987        & 11987             & 4177              \\ \cline{2-5} 
\multicolumn{1}{|l|}{}                                & \textit{Min}     & 10            & 10                 & 10                 \\ \cline{2-5}
\multicolumn{1}{|l|}{}                                & \textit{Average} & 112           & 152               & 60                \\ \hline
\multicolumn{5}{|l|}{\textbf{}}                                     \\ \hline
\multicolumn{1}{|l|}{\multirow{3}{*}{\textbf{Density}}} & \textit{Max}     & 0.2167        & 0.1727             & 0.2167              \\ \cline{2-5} 
\multicolumn{1}{|l|}{}                                & \textit{Min}     & 0.0001            & 0.0003                 & 0.0001                 \\ \cline{2-5}
\multicolumn{1}{|l|}{}                                & \textit{Average} & 0.0361           & 0.0440               & 0.0301                \\ \hline

\end{tabular}
\caption{Stats of Big-Vul dataset}
\label{tab:big-vul}
\end{table}

\subsection{Results}
\label{results}

We present the classification results of our evaluation on both datasets in Table~\ref{tab:results}. The table compares the performance of five models across both datasets as well as between SEGNN and the baselines. The results indicate that the proposed dataset, CVEFGE, improves the performance of models by up to 20\% when compared to Big-Vul. Similarly, the SEGNN framework outperformed all baselines on both datasets and achieved the maximum performance compared to baselines. These results clearly demonstrate the expressiveness of our proposed dataset in vulnerability detection models. We also show the training and testing curves of SEGNN on both datasets in Figure~\ref{fig:curves}. On both datasets, we can observe a significant difference among the results. These findings clearly show the superiority of the proposed dataset over the existing Big-Vul dataset.


\begin{table}[!t]
    \centering
    \begin{tabular}{|c|c|c|}  
    \hline
         \textbf{Model}& \textbf{Big-Vul Acc ($\%$)} & \textbf{CVEFGE Acc ($\%$)} \\ \hline
        MEWISPool \cite{nouranizadeh2021maximum}    & {\color{red}$63.0$} & $61.0$ \\ \hline
        GraphSAGE \cite{hamilton2017inductive}    & $58.8$ &  {\color{blue}$78.22$}\\ \hline
        GAT \cite{velivckovic2017graph}          & $63.3$ & {\color{blue}$75.69$} \\ \hline
        DGSD \cite{said2021dgsd}        &     $62.2 $  & {\color{blue}$79.1$}\\ \hline
        \textbf{SEGNN}& \textbf{{\color{blue}$65.4$}} & \textbf{{\color{blue}$79.47$}}\\ \hline
    \end{tabular}
    \caption{A comparison of the proposed frameworks, SEGNN against existing methods on two datasets. {\color{blue} blue} indicates the improved results in comparison to Big-Vul, and the baselines.} 
    \label{tab:results}

\end{table}



\section{Conclusion}

This work offers two main contributions. First, the collection of a properly curated C/C++ source code vulnerability dataset, CVEFGE, which contains a high level of source code complexity. CVEFGE contains a rich set of open-source functions as well as their control flow graphs and learned representations, which simplifies the machine learning task to a large extent. It is crawled from the CVE database, which contains authentic vulnerabilities along with their fixed versions. The second contribution of this work is introducing a learning framework that is based on graph neural networks to learn code's semantic and structural representations. It consists of a gated graph convolution layer with attention mechanism, pooling, and a fully connected layer. We evaluate the proposed dataset and model against four baselines and one existing dataset. Our results demonstrate state-of-the-art performance on vulnerability identification task.  

\bibliographystyle{ACM-Reference-Format}
\bibliography{references.bib}


\end{document}